\documentstyle[aps,prd,epsfig,twocolumn]{revtex}
\draft

\begin{document}
\title{A Cosmological Three Level Neutrino Laser}
\author{Steen Hannestad and Jes Madsen}
\address{Theoretical Astrophysics Center and
Institute of Physics and Astronomy,
University of Aarhus, 
DK-8000 \AA rhus C, Denmark}
\date{\today}
\maketitle

\begin{abstract}
We present a calculation of a neutrino decay scenario in the
early Universe. The specific decay is $\nu_{2} \rightarrow
\nu_{1} + \phi$, where $\phi$ is a boson.
If there is a neutrino mass hierarchy, $m_{\nu_{e}} < m_{\nu_{\mu}} 
< m_{\nu_{\tau}}$,
we show that
it is possible to generate stimulated decay and effects similar
to atomic lasing without invoking new neutrinos, 
even starting from identical neutrino distributions. 
Under the right
circumstances the decay can be to very low momentum boson states
thereby producing something similar to a Bose condensate, 
with possible consequences for structure formation.
Finally, we argue that this type of
decay may also be important other places in early Universe physics.
\end{abstract}

\pacs{95.35, 13.35.H, 98.62.L}

\section{Introduction}
\label{sec:Intro}

Any neutrino that is absolutely stable must have a
mass less than roughly 30 eV in order not to overclose the 
Universe \cite{kolb}. On the other hand, it is very plausible 
that neutrinos have a non-zero mass. This makes neutrino decays 
necessary if their mass is larger than the cosmological limit
for stable neutrinos.
There are several possible decay modes \cite{kim}, 
for example the simple
radiative decay $\nu_{2} \rightarrow \nu_{1} + \gamma$ or the 
flavor changing neutral current weak
decay $\nu_{2} \rightarrow \overline{\nu}_{1} \nu_{1}
\nu_{1}$. Another possibility is the decay
\begin{equation}
H \rightarrow F + \phi,
\end{equation}
where $H$ is a heavy neutrino, $F$ is a light neutrino and $\phi$ is 
a light scalar particle (for example the majoron). It is not
essential that the boson should have spin 0, but it simplifies 
calculations significantly.
All of these decay modes have been investigated in an early Universe
context \cite{steigman}.
We choose to focus on the last decay type in order to investigate some
very interesting non-equilibrium effects that pertain to this decay.
It turns out that 
under the right circumstances the heavy neutrino will decay
by stimulated decay into $F$ and $\phi$ instead of just decaying
thermally. This runaway process leads to the formation of a
very low momentum component of the scalar particle, $\phi$,
while the rest are in an almost thermal distribution.
The scenario has been described by Madsen \cite{madsen} and
subsequently in more detail by Kaiser,
Malaney and Starkman \cite{kaiser,starkman}, 
the idea being that the $\phi$ particle could
make up the dark matter in our Universe. 
One of the currently favoured models of structure formation is the 
so-called Mixed Dark Matter (MDM) model \cite{klypin}. 
This is essentially a
Cold Dark Matter (CDM) model, but with a small matter content of 
Hot Dark Matter (HDM) that lessen the fluctuations on small scales.
The $\phi$ particles could fit both roles, the cold component
acts as cold dark matter and the thermal component as hot dark
matter.

There are, however, several problems with this scenario. First of all,
the favored mixed dark matter models consist of roughly 5\% baryons,
25\% HDM and 70\% CDM. If there is too much HDM, structure formation
will be affected so as to produce too little clustering on small 
scales (if normalised to COBE data). The simple scenario of
Refs.\ \cite{madsen,kaiser,starkman} predicts only about 35\% CDM
and 60\% HDM, which is incompatible with the observed structure
\cite{larsen}.

We develop further this model and 
in Section 2 we list the relevant equations for calculating the
particle distribution functions resulting from the decay.
In Section 3 we describe a new way of producing neutrino lasing
that is very similar to an atomic 3-level laser.
Here it is possible to achieve lasing without invoking new neutrinos
and to increase the CDM fraction, which however still remains well
below 70\%.
Section 4 contains a description of the numerical results from
our investigation of this 3-level model and, finally, Section 5
contains a discussion of the possible consequences of this type
of decays as well as a discussion of various astrophysical limits
on them. From this discussion it will become clear that lasing decays
are most likely to take place after the epoch of nucleosynthesis
($T \simeq 1 - 0.1 \text{MeV}$). Of course they are also required
to take place before matter-radiation equality ($T \simeq 1 \text{eV}$)
if the bosons are to act as ordinary MDM.

\section{The Basic Scenario}

The fundamental equation that describes the evolution of all particle
species during the decay is the Boltzmann equation, 
\begin{equation}
\text{L}[f] =
C_{\mbox{\scriptsize{dec}}}[f] + C_{\text{coll}}[f].
\label{boltzmann2}
\end{equation}
The left hand side is the basic Liouville operator in an expanding 
Universe
\begin{equation}
\text{L}[f] = \frac{\partial f}{\partial t}-\frac{{d} R}
{{d} t}\frac{1}{R}p\frac{\partial f}{\partial p}.
\end{equation}
The right hand side of Eq.\ (\ref{boltzmann2})
involves the specific physics of the
decay process as well as possible other collision processes such as
elastic scattering and annihilation. We shall assume that the term
$C_{\text{coll}}$ can be neglected, but possible limitations to this
assumption will be discussed later.
For simplicity we assume that the
decay is isotropic in the rest-frame (see, however, Refs. 
\cite{kaiser,starkman}
for a discussion of the case of anisotropic decays).

The decay terms for the different particles have been calculated
in for example Refs. \cite{kaiser,kawasaki}. They are of the form

\begin{equation}
C_{\mbox{\scriptsize{dec}}}[f_{H}] = 
- \frac{m_{H}^{2}}{\tau m_{0} E_{H}
p_{H}}
\int_{E_{\phi}^{-}}^{{E_{\phi}^{+}}}dE_{\phi}
\Lambda(f_{H},f_{F},f_{\phi}),
\end{equation}

\begin{equation}
C_{\mbox{\scriptsize{dec}}}[f_{F}] = 
\frac{g_{H}}{g_{F}} \frac{m_{H}^{2}}{\tau m_{0} E_{F} 
p_{F}}
\int_{E_{H}^{-}}^{{E_{H}^{+}}}dE_{H}
\Lambda(f_{H},f_{F},f_{\phi}),
\end{equation}

\begin{equation}
C_{\mbox{\scriptsize{dec}}}[f_{\phi}] = 
\frac{g_{H}}{g_{\phi}}
\frac{m_{H}^{2}}{\tau m_{0} E_{\phi} 
p_{\phi}}
\int_{E_{H}^{-}}^{{E_{H}^{+}}}dE_{H}
\Lambda(f_{H},f_{F},f_{\phi}),
\label{bosdec}
\end{equation}
where
$\Lambda(f_{H},f_{F},f_{\phi}) = f_{H}(1-f_{F})(1+f_{\phi})-
f_{F}f_{\phi}(1-f_{H})$, 
$m_{0}^{2} = m_{H}^{2}-2(m_{\phi}^{2}+m_{F}^{2})+
(m_{\phi}^{2}-m_{F}^{2})^{2}/m_{H}^{2}$.
$\tau$ is the lifetime of the heavy neutrino and $g$ is the
statistical weight of a given particle. We use $g_{H} = g_{F} = 2$ and
$g_{\phi} = 1$, corresponding to $\phi = \overline{\phi}$ 
\footnote{All calculations may be redone using $g_{\phi} = 2$.
The changes are not very significant and tend to lessen the
lasing effect described below \cite{madsen,kaiser}.}.
The integration limits are
\begin{eqnarray}
E_{H}^{\pm} (E_{i}) & = & \frac{m_{0}m_{H}}
{2m_{i}^{2}}[E_{i}(1+4(m_{i}/m_{0})^{2})^{1/2} \pm \\ \nonumber
& & (E_{i}^{2}-m_{i}^{2})^{1/2}],
\end{eqnarray}
and
\begin{equation}
E_{i}^{\pm} (E_{H}) = \frac{m_{0}}{2m_{H}}
[E_{H} (1+4(m_{i}/m_{0})^{2})^{1/2} \pm p_{H}],
\label{energyi}
\end{equation}
where the index $i = F,\phi$.

For non-relativistic decays, $E_{\phi} \simeq m_{H}/2 \gg T$.
This means that the resulting fraction of low momentum bosons is
vanishing because they are kinematically inaccessible. 
However, for relativistic decays this is not the case. Here the
minimum boson energy is roughly 
$E_{\phi}^{-} \simeq \frac{1}{2}p_{H}[2 m_{\phi}^{2}/m_{H}^{2} + 
\frac{1}{2} m_{H}^{2}/p_{H}^{2}]
\ll E_{H}$ which can be seen from Eq.\ (\ref{energyi}).
This corresponds to bosons being emitted
backwards relative to the direction of motion of the 
parent particle. In relativistic decays
it is thus possible to access very low energy boson states. 
This is essential for the phenomenon described below.

Something very interesting can happen if there is initially an
overabundance of $H$ over $F$. To see this, let us look at the 
structure of the decay term for the boson
Eq.\ (\ref{bosdec}). Following Ref.\ \cite{kaiser}
we make the following rough approximation, $1+f_{\phi} \simeq f_{\phi}$.
The decay term then reduces to
\begin{equation}
C_{\mbox{\scriptsize{dec}}}[f_{\phi}] = 
\frac{2 m_{H}^{2}}{\tau m_{0} E_{\phi} p_{\phi}} f_{\phi}
\int_{E_{H}^{-}}^{{E_{H}^{+}}}dE_{H} [f_{H}-f_{F}].
\label{decstruc}
\end{equation}
If $f_{H} > f_{F}$ this indicates an exponential growth of
$f_{\phi}$. 
This effect has been called neutrino lasing because
it is very similar to normal lasing, the overabundance of
$H$ over $F$ corresponding to a population inversion. Exactly as
in a laser, the process saturates when $f_{H} = f_{F}$. The 
remaining neutrinos will then decay to an almost thermal $\phi$
distribution as they go non-relativistic
\footnote{Note that as mentioned we have only considered 
isotropic decays, but as
mentioned in Ref.\ \cite{kaiser} even if the decay is fully
polarized and the bosons are emitted backwards, there will
only be a rather small increase in the number of low momentum
bosons. If the bosons are preferentially emitted in the forward
direction there will be no lasing. Refs.\ \cite{kaiser,starkman} have
good discussions of these polarisation effects.}.

Note that in all our actual calculations we use 
a zero initial abundance of bosons
before the decay commences. This is not essential for the lasing
phenomenon, but the final fraction of cold bosons is decreased
if there is an initial thermal population
(In Ref. \cite{kaiser} it was assumed that the bosons decoupled
prior to the QCD phase transition so that their number density 
was depleted by a factor of 8 relative to the other species).

The effect will be larger for low momentum
states because of the $(E_{\phi}p_{\phi})^{-1}$ term. This means that
if the low momentum states are accessible, the bosons will
cluster in these states and produce something similar to a Bose
condensate. For this to happen it is a necessary condition that
the 
decay is relativistic, because otherwise no low momentum states
can be accessed.

This phenomenon was first investigated in 
Refs.\ \cite{madsen,kaiser,starkman}.  
However, Ref.\ \cite{madsen} used a somewhat different approach, namely
equilibrium thermodynamics,
where the decays can lead to the formation of a
true Bose condensate. In the above scenario, following 
Refs.\ \cite{kaiser,starkman}, no
assumption of equilibrium has to be made.

\section{The 3-Level Laser}

It was assumed in Refs.\ \cite{kaiser,starkman} that $F$ is a 
hitherto unknown neutrino that decouples prior to the QCD
phase transition (for example a right handed component of one of
the known neutrinos).
This is the only way of achieving lasing in the above 2-level
decay. However, it is actually possible to obtain lasing
with only the three known neutrinos. The effect is somewhat
similar to an atomic 3-level laser, where particles are pumped from
the ground state to the top level. From there they decay
rapidly to the lasing state and from the lasing state they
decay by stimulated decay to the ground state.
Exactly the same situation can occur for neutrinos if we have a
three generation mass hierarchy, $m_{\nu_{e}} < m_{\nu_{\mu}} 
< m_{\nu_{\tau}}$ (the stable particles $\nu_{e}$ and $\phi$ are
of course required to have masses below the cosmological limit
for stable particles, $m < 30 \text{eV}$, and can therefore
be considered essentially massless in the calculations).
Let us assume that we have the same generic decay mode as before.
The possible majoron emitting decays are then 
\footnote{We choose to focus alone on the majoron decays. In principle
there might be other relevant decay modes as mentioned previously.}
\begin{eqnarray}
\nu_{\tau} \rightarrow \nu_{\mu} + \phi \label{taumu} \\
\nu_{\tau} \rightarrow \nu_{e} + \phi \label{taue} \\
\nu_{\mu} \rightarrow \nu_{e} + \phi.
\end{eqnarray}
We shall assume that the second decay mode, Eq.\ (\ref{taue}),
 is suppressed so that it is negligible relative to  
the first, Eq.\ (\ref{taumu}). This assumption is not ruled out
by the physical models described later.

If the decays are non-relativistic, nothing exciting happens
and the decays are thermal. However, if all the decays are 
relativistic and the free decay of $\nu_{\tau}$ proceeds
much faster than that of $\nu_{\mu}$, then we can
achieve lasing.
The $\tau$ neutrino now functions as the top level in a normal
laser, it feeds particles to the lasing state, $\nu_{\mu}$.
Particles in this state then decay to the ground state, 
$\nu_{e}$, by stimulated decay.
The condition for the lasing process to work is that $f_{\nu_{\mu}} >
f_{\nu_{e}}$, but because of the fast decay of the $\tau$ neutrino
this is automatically fulfilled for a large 
temperature range between $m_{\nu_{\tau}}$ and $m_{\nu_{\mu}}$.

Let us for the moment assume that the 
$\nu_{\tau}$ decays completely before
$\nu_{\mu}$ begins to decay. 
It is then easy to see why the amount of lasing increases if
$\nu_{\tau}$ has decayed relativistically.
We now start with a
non-thermal distribution of muon neutrinos. If $\nu_{\tau}$
decays while still relativistic, the distribution of $\nu_{\mu}$ 
will be significantly colder than thermal.
If we again look at the structure of Eq.\ (\ref{decstruc}), we see
that for $n_{H}$ given, the
integral increases if the distribution of $H$ gets colder
because of the $p^{2}$ factor in the number density integral. This
means that the relativistic decay to muon neutrinos enhances the
lasing from $\nu_{\mu}$ to $\nu_{e}$. The effect is opposite if
the $\nu_{\tau}$ decay is non-relativistic, and the lasing diminishes
or disappears completely.

% --------------------------------------
% Frontpage figure
% --------------------------------------
\begin{figure}[h]
\begin{center}
\epsfysize=6.5truecm\epsfbox{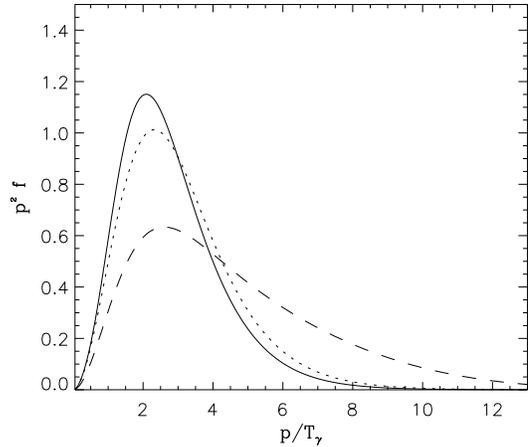}
\vspace{0truecm}
\end{center}
\baselineskip 17pt
\caption{The distribution function of the light neutrino, $F$, 
after complete decay of the heavy neutrino, $H$, for different
lifetimes of the heavy neutrino using specifically 
$m_{H} = 2 \text{MeV}$, $m_{F} = 0.2 \text{MeV}$
and $m_{\phi} = 0.005 \text{MeV}$. The full curve is for $\tau_{H} = 0.2 
\text{s}$, the dotted for $\tau_{H} = 1 \text{s}$ and the
dashed for $\tau_{H} = 20 \text{s}$.}
\label{fig1}

\end{figure}

Fig.\ \ref{fig1} shows the momentum distribution of the light 
neutrino after
complete decay of the heavy neutrino. We see exactly that the
light neutrino distribution gets colder if the decay is relativistic
and hotter if it is non-relativistic.
Of course, it is always a necessary condition for
lasing to occur that $\nu_{\mu}$ decays
relativistically because otherwise it cannot populate low momentum
boson states at all.

\section{numerical results}

We have investigated numerically the behavior of this 3-level
decay for different parameters by solving explicitly the 
coupled Boltzmann equations. Our numerical routine uses a grid
in comoving momentum space and evolves forward in time using
a Runge-Kutta integrator. The time evolution of the scale factor, $R$,
and the photon temperature, $T_{\gamma}$, are found by use
of the energy conservation equation
${d}(\rho R^{3})/dt + p {d}(R^{3})/dt = 0$ and the Friedmann
equation $H^{2} = 8 \pi G \rho/3$.

% --------------------------------------
% Frontpage figure
% --------------------------------------
\begin{figure}[h]
\begin{center}
\epsfysize=6.5truecm\epsfbox{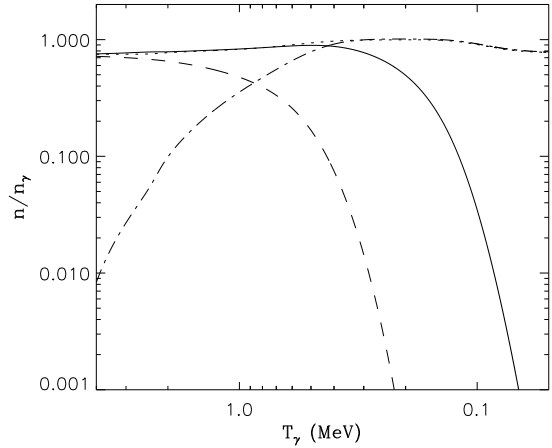}
\vspace{0truecm}
\end{center}
\baselineskip 17pt
\caption{The evolution of number density for the different involved
particles. The full curve is $\nu_{\mu}$, the dotted $\nu_{e}$,
the dashed $\nu_{\tau}$ and the dot-dashed $\phi$. The parameters
used are $m_{\nu_{\tau}} = 2 \text{MeV}$, $m_{\nu_{\mu}} = 0.5 \text{MeV}$,
$m_{\phi} = m_{\nu_{e}} = 0.005 \text{MeV}$ and 
$\tau_{\nu_{\tau}} = 0.2 \text{s}$, $\tau_{\nu_{\mu}} = 1.0 \text{s}$.}
\label{fig2}

\end{figure}

In Fig.\ \ref{fig2} we show an example of the number density 
evolution of
the different involved particles. It is seen that in a large 
temperature interval, the $\nu_{\mu}$ number density is 
slightly higher than
that of $\nu_{e}$. This leads to a very fast growth of the boson
number density and is identified with the region where lasing
takes place. The final number density of $\phi$ and $\nu_{e}$
are the same because of neutrino number conservation
(neglecting a possible small initial abundance of $\phi$)
because even though the $\nu_{e}$ abundance starts out higher,
three $\phi$'s are produced for every two $\nu_{e}$.

Since both $\nu_{e}$ and $\phi$ are almost massless there are 
essentially four free parameters in our equations. The first is the
mass scale, defined for example by the $\tau$ neutrino mass.
This mass scale is unimportant for the shape of the spectra
(this is only true if the decays take place below the weak decoupling
temperature
$T_{\text{dec}} \simeq 2.3 \text{MeV}$. If the free decay temperature
is larger than $T_{\text{dec}}$ the $\phi$ distribution will quickly
thermalize. Only well below $T_{\text{dec}}$ can the non-equilibrium
lasing proceed, but now with an initial thermal $\phi$ distribution.
In our numerical work we choose that the free decay 
temperature to be lower than the decoupling temperature). 
% --------------------------------------
% Frontpage figure
% --------------------------------------
\begin{figure}[h]
\begin{center}
\epsfysize=6.5truecm\epsfbox{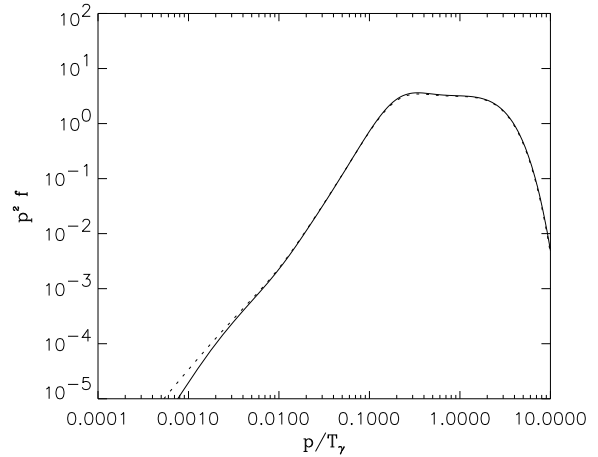}
\vspace{0truecm}
\end{center}
\baselineskip 17pt
\caption{Final boson distribution for two different decay models. The
full curve is for the parameters
$m_{\nu_{\tau}} = 2 \text{MeV}$, $m_{\nu_{\mu}} = 0.5 \text{MeV}$,
$m_{\phi} = m_{\nu_{e}} = 0.005 \text{MeV}$ and 
$\tau_{\nu_{\tau}} = 1.125 \text{s}$, $\tau_{\nu_{\mu}} = 18 \text{s}$.
The dashed curve is for
$m_{\nu_{\tau}} = 1 \text{MeV}$, $m_{\nu_{\mu}} = 0.25 \text{MeV}$,
$m_{\phi} = m_{\nu_{e}} = 0.0025 \text{MeV}$ and 
$\tau_{\nu_{\tau}} = 4.5 \text{s}$, $\tau_{\nu_{\mu}} = 72 \text{s}$.
Note that although we have chosen values for $m_{\nu_{e}}$ and $\phi$
which is orders of magnitude larger than the 
experimental/cosmological limit, this 
does not influence the results as long as the values are small compared to
all other relevant masses and temperatures.}
\label{fig3}

\end{figure}
In Fig.\ \ref{fig3} we show an example
of changing the normalisation ($m_{\nu_{\tau}}$)
 by a factor of two. This leads to
an almost identical final boson distribution.
The other three parameters are
\begin{eqnarray}
\alpha_{\nu_{\tau}} & \equiv & \left(\frac{m_{\nu_{\tau}}}
{3 \text{MeV}}\right)^{2} \left(\frac{\tau_{\nu_{\tau}}}{1 \text{s}}
\right) \\
\alpha_{\nu_{\mu}} & \equiv & \left(\frac{m_{\nu_{\mu}}}
{3 \text{MeV}}\right)^{2} \left(\frac{\tau_{\nu_{\mu}}}{1 \text{s}}
\right) \\
\beta & \equiv & \frac{m_{\nu_{\mu}}}{m_{\nu_{\tau}}}.
\end{eqnarray}
The first two parameters describe how relativistic the free decays
of the particles are. A thermal particle shifts from relativistic to 
non-relativistic at a temperature of roughly $T \simeq m/3$.
When the Universe is radiation dominated
\begin{equation}
\frac{t}{1 \text{s}} \simeq \left(\frac{T}{1 \text{MeV}}\right)^{-2}.
\end{equation}
Therefore if the decay is relativistic 
\begin{equation}
\tau < t(T = m/3) \simeq \left(\frac{m}{3 \text{MeV}}\right)^{-2} 
\text{s}.
\end{equation}
Thus, if $\alpha_{i} < 1$ the decay is relativistic, whereas if 
$\alpha_{i} > 1$ it is non-relativistic.
The third parameter is just the dimensionless $\nu_{\mu}$ mass.

% --------------------------------------
% Frontpage figure
% --------------------------------------
\begin{figure}[h]
\begin{center}
\epsfysize=6.5truecm\epsfbox{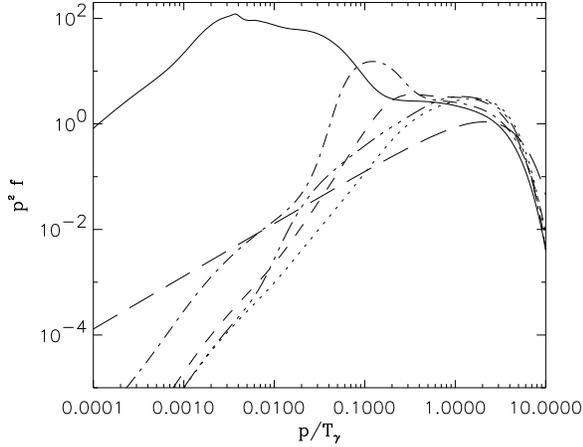}
\vspace{0truecm}
\end{center}
\baselineskip 17pt
\caption{The momentum spectrum of $\phi$ after complete decay 
for different decay parameters. 
The full curve is $\alpha_{\nu_{\tau}} = 0.09$, 
$\alpha_{\nu_{\mu}} = 0.01$ and $\beta = 0.15$.
The dotted is for $\alpha_{\nu_{\tau}} = 1$, 
$\alpha_{\nu_{\mu}} = 1$ and $\beta = 0.25$.
The dashed is for $\alpha_{\nu_{\tau}} = 0.5$, 
$\alpha_{\nu_{\mu}} = 0.5$ and $\beta = 0.25$.
The dot-dashed is for $\alpha_{\nu_{\tau}} = 1$, 
$\alpha_{\nu_{\mu}} = 0.03$ and $\beta = 0.25$.
The dot-dot-dot-dashed is for $\alpha_{\nu_{\tau}} = 0.09$, 
$\alpha_{\nu_{\mu}} = 1$ and $\beta = 0.25$.
The long-dashed is a thermal Bose distribution with the same
number density.}
\label{fig4}

\end{figure}

Fig.\ \ref{fig4} shows the final boson distribution after 
complete decay for 
several different choices of decay parameters
(for comparison we have also shown a thermal Bose 
distribution with the same number density). It is seen
that in order to get a large fraction of cold particles, we need to have
$\alpha_{\nu_{\tau}} \ll 1$ and $\alpha_{\nu_{\mu}} \ll 1$. 
The reason for this was described in the previous section.
% --------------------------------------
% Frontpage figure
% --------------------------------------
\begin{figure}[h]
\begin{center}
\epsfysize=6.5truecm\epsfbox{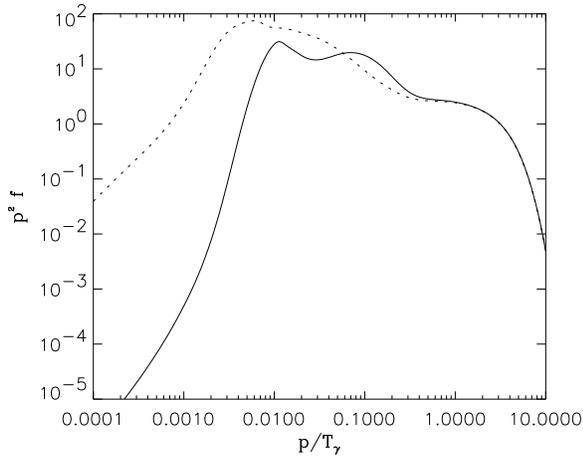}
\vspace{0truecm}
\end{center}
\baselineskip 17pt
\caption{Final momentum distribution of $\phi$. The full curve is for
$\alpha_{\nu_{\tau}} = 0.09$, 
$\alpha_{\nu_{\mu}} = 0.03$ and $\beta = 0.25$.
The dotted curve is for $\alpha_{\nu_{\tau}} = 0.09$, 
$\alpha_{\nu_{\mu}} = 0.03$ and $\beta = 0.15$.}
\label{fig5}

\end{figure}
Fig.\ \ref{fig5} shows the effect of changing 
the relative mass of the two heavy neutrinos.
We see here that to get a large cold fraction we also need to have 
$m_{\nu_{\mu}}/m_{\nu_{\tau}} \ll 1$.
This is also understandable if we look at the allowed momentum range
of the produced bosons from Eq.\ (\ref{energyi})
\begin{equation}
\frac{p_{\phi}^{-}}{T_{\gamma}} = \frac{1}{2}(1-\beta^{2})
\left[\sqrt{\frac{p_{\nu_{\tau}}^{2}}{T_{\gamma}^{2}}+
\frac{m_{\nu_{\tau}}^{2}}{T_{\gamma}^{2}}} - 
\frac{p_{\nu_{\tau}}}{T_{\gamma}}\right].
\label{energyrange}
\end{equation}
If $\beta$ is close to one we can get 
very cold bosons. This means that the bosons will tend to cluster
in low momentum states, but because of energy conservation we get
a very hot $\nu_{\mu}$ distribution.
On the other hand, if $\beta \rightarrow 0$,
then the lowest boson states are unaccessible and we get instead a
colder $\nu_{\mu}$ distribution. 
This in turn leads to more lasing in the lasing state
and even though the fraction of cold bosons coming from the initial
$\nu_{\tau}$ decay gets smaller, this is more than compensated for
by the larger fraction coming from the $\nu_{\mu}$ decay.
This is indeed why the cold fraction increases when $\beta$ 
decreases.

\begin{figure}[h]
\begin{center}
\epsfysize=6.5truecm\epsfbox{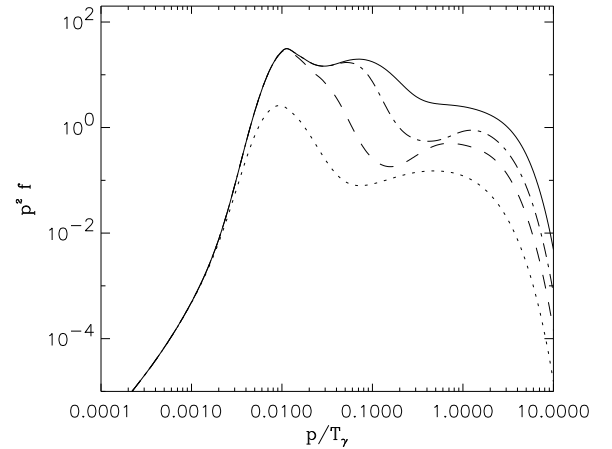}
\vspace{0truecm}
\end{center}
\baselineskip 17pt
\caption{The temperature evolution of the $\phi$-distribution.
The decay parameters are $\alpha_{\nu_{\tau}} = 0.09$, 
$\alpha_{\nu_{\mu}} = 0.03$ and $\beta = 0.25$.
The dotted curve is at $T_{\gamma} = 2.3 \text{MeV}$, the
dashed at $T_{\gamma} = 1.1 \text{MeV}$ and the dot-dashed at
$T_{\gamma} = 0.5 \text{MeV}$. The full curve is the
final distribution after complete decay.} 
\label{fig6}

\end{figure}

Thus, if we want to study the region of parameter space with 
possible interest
for structure formation scenarios we have essentially limited ourselves
to the region of ultrarelativistic decays with a large mass difference
between the two heavy neutrinos.
It is also interesting to follow the evolution of the boson distributions
with temperature to see how the different states populate.
Fig.\ \ref{fig6} shows the temperature evolution of the boson
distribution for the decay parameters
$\alpha_{\nu_{\tau}} = 0.09$, $\alpha_{\nu_{\mu}} = 0.03$ and 
$\beta = 0.25$.

% --------------------------------------
% Frontpage figure
% --------------------------------------

\begin{figure}[h]
\begin{center}
\epsfysize=6.5truecm\epsfbox{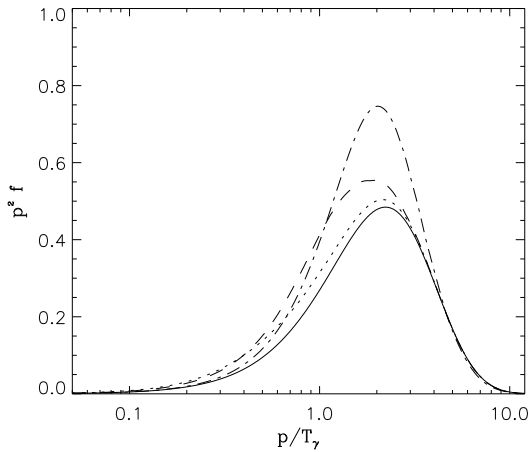}
\vspace{0truecm}
\end{center}
\baselineskip 17pt
\caption{The temperature evolution of the $\nu_{\mu}$ distribution.
The decay parameters are as in Fig.\ \protect\ref{fig6}.
The full curve is the initial thermal $\nu_{\mu}$ distribution.
The dotted curve is at $T_{\gamma} = 2.3 \text{MeV}$, the dashed at
$T_{\gamma} = 1.1 \text{MeV}$ and the dot-dashed curve is 
at $T_{\gamma} = 0.5 \text{MeV}$.}
\label{fig7}

\end{figure}

The minimum possible boson energy at a given temperature is given by
Eq.\ (\ref{energyrange}),
where instead of $\nu_{\tau}$ we use $\nu_{\mu}$ and for $\beta$ we
use 0.
An approximate lower limit can be calculated by setting 
${p_{\nu_{\mu}}}/{T_{\gamma}} = 3$, corresponding roughly 
to the mean value from a normal thermal distribution. 
This lower limit is an increasing function with decreasing temperature
and indicates roughly the momentum where the distribution ``decouples''
(decoupling here means that the states are kinematically inaccessible)
at a given temperature.
For the model shown in Fig.\ \ref{fig6} we get
$p_{\phi}^{-}/T_{\gamma} = \text{0.0039, 0.017, 0.081}$ for 
photon temperatures of 2.3, 1.1 and 0.5 MeV.
We see that the lower limit at a given temperature corresponds roughly to
the momentum where the distribution decouples. Of course the lower
limit will always be somewhat higher than the actual decoupling point,
but it helps to understand the evolution of the spectra.
Fig.\ \ref{fig7} shows the corresponding evolution of $\nu_{\mu}$. 
We see the before mentioned effect of $\nu_{\tau}$ decay producing
a cold $\nu_{\mu}$ distribution and thereby enhancing lasing.

% --------------------------------------
% Frontpage figure
% --------------------------------------

However, even in the case where the decays are ultrarelativistic
and the mass ratio small, the fraction of cold bosons is only 
moderate. For example in the case where $\alpha_{\nu_{\tau}} = 0.09$,
$\alpha_{\nu_{\mu}} = 0.01$ and $m_{\nu_{\mu}}/m_{\nu_{\tau}} = 0.15$,
the fraction of bosons with ${p_{H}}/{T_{\gamma}} < 0.3$ is only
44\%. If the boson was to act as the dark matter in our Universe, a
much larger fraction would be needed.
Note, however, that even if the obtainable cold fraction is too
small, it is comparable with or larger than that obtained in 
the model of Refs.\ \cite{madsen,kaiser,starkman}. 
Furthermore, it is achieved without invoking an unknown neutrino state.

\section{discussion}

We have developed further the scenario originally proposed in
Refs.\ \cite{madsen,kaiser,starkman} 
by introducing a physically different way of obtaining
neutrino lasing. However, this model still suffers from one of the same 
shortcomings that the original model did, namely that it is 
impossible to achieve sufficiently large fractions of cold bosons
for structure formation.

An interesting consequence of the 3-level decay is that
the number density of electron neutrinos is increased by a factor
of almost three relative to the standard case. If the decay takes 
place before or during Big Bang Nucleosynthesis (BBN), 
then it can significantly change the outcome
because the electron neutrinos are very important for the neutron
to proton conversion reactions.
The higher than usual number of electron neutrinos will tend to keep
equilibrium longer than in the standard case, meaning that less
neutrons survive to form helium. The helium abundance will then 
be lower than in the standard case. Furthermore, the ratio
of low energy electron neutrinos is significantly higher than
normal. This favors the neutron to proton reaction because
of the mass difference between the neutron and the proton
and leads to the opposite effect of the one described by Gyuk and Turner
\cite{gyuk}, 
where a massive $\tau$ neutrino decays to electron neutrinos,
thereby enhancing the neutron fraction and allowing a
very high baryon density without violating the BBN constraints.
However, in their scenario, the decaying neutrino is non-relativistic
and the resulting electron neutrinos therefore have very high 
energy (the opposite of our case).

Decays producing lasing effects are, however, not likely to take
place during BBN. The reason is that neutrinos are kept in equilibrium
by the weak reaction until roughly $T \simeq 2 \text{MeV}$. 
All neutrino distributions are then of equilibrium form 
and no lasing will take place.
If the mass of any of the heavy neutrinos is of the same order
or larger than this temperature, we will essentially
have a non-relativistic decay after decoupling, but this will not
produce lasing. Thus, the heaviest neutrino in our scenario 
must be substantially lighter than 2 MeV, thus making it improbable
that it should decay during BBN. In general neutrino decays affecting
BBN may well take place, but then they will not produce lasing
effects. Therefore we have not treated this case in the present paper.

If we turn to more specific physical models, the most natural
candidate for the light scalar particle, $\phi$, is the majoron.
There exist several majoron models, all of them designed to
explain the very small neutrino masses compared to the other fermions.
In the simplest models, the decay $\nu_{2} \rightarrow \nu_{1} 
+ \phi$ is forbidden at tree level \cite{kim}, making the
lifetime too long to be of interest. 
 However, there are more complicated models where
the lifetime can be short.
In the simplest case the majoron couples to the neutrino via
a Yukawa interaction
\begin{equation}
\newfont{\sss}{cmsy10}
\text{\begin{sss} L \end{sss}} = i \frac{g}{2} J 
\overline{\nu^{c}}_{1} \gamma_{5} \nu_{2} + \text{h.c.} 
\end{equation}
From this generic interaction several limits to the coupling 
strength, $g$, and therefore to the lifetime can be derived.
For example neutrinoless double $\beta$ decays give an upper limit
on $g$ of roughly $10^{-4}$ \cite{berezhiani}. Other limits can
be derived from the neutrino signal of SN1987A
\cite{raffelt}, but as mentioned
in Ref.\ \cite{raffelt}, it is very difficult to get a clear 
picture of the current limits. A coupling constant of $10^{-4}$ 
gives a restframe decay lifetime of \cite{kim} 
\begin{equation}
\tau = \frac{32 \pi}{g^{2} m_{2}} > 10^{-5} \text{s} \times
\left(\frac{m_{2}}{1 \text{eV}}\right)^{-1}. 
\end{equation}
This limit is rather
weak and leaves open a large region of parameter space for
relativistic decays in the early Universe.
Note that for our model to work the decay $\nu_{\tau} \rightarrow
\nu_{e}$ must be suppressed. This corresponds to the coupling
$g_{\tau e}$ being very small compared to the other coupling
constants.

Finally, we have assumed that the scattering and annihilation 
reactions between neutrinos and majorons are weak compared with
the decay reaction. If, as we have assumed, the decay is not
forbidden at tree level this is the case because the scattering
and annihilations all are of order $g^{4}$, whereas the decays
are of order $g^{2}$. Were the annihilation and scattering reactions
to be important we would not obtain any cold bosons because the
low momentum states would not be decoupled. Therefore, it is
absolutely necessary that the decays are stronger than the
scattering and annihilation reactions. 
Note, that even if the decay is of order $g^{2}$ and the scattering
and annihilation $g^{4}$ the decay rate will be damped by a
factor $\gamma^{-1} = m/E$ which can be very large if the decay is
extremely relativistic. Thus, in the limit of 
$\gamma \rightarrow \infty$ we have to account for
scatterings and annihilations, but $\gamma$ would have to be extremely
large because of the very small value of $g$.

As far as the neutrino masses are concerned, they are naturally
explained by see-saw models. Then only the lefthanded components
participate in the interactions because of the very large mass
of the right handed components. The possible Majoron decays are then
exactly
\begin{eqnarray}
\nu_{i}(L) & \rightarrow & \nu_{j}(L) + \phi \\
\nu_{i}(L) & \rightarrow & \overline{\nu}_{j}(R) + \phi
\end{eqnarray}
as required in our lasing model.

In conclusion, what we have shown is that it is possible to generate
large non-equilibrium effects, even starting from identical neutrino
distributions. Very few ``fine-tunings'' are necessary in order to 
achieve lasing.
Thus, these effects may be present many
places in early Universe physics, both because the decay 
types are generic and because the early Universe is an 
inherently non-equilibrium environment.
This was also noted in Ref.\ \cite{starkman}, but our specific
model shows that the relevant range of models is much larger than
originally thought. It is therefore of considerable interest to search
for other physical scenarios where the lasing process 
(and other similar non-equilibrium processes) may be effective.

\acknowledgements
We thank Georg Raffelt for valuable comments on a draft version of
this paper.
This work was supported by the Theoretical Astrophysics Center
under the Danish National Research Foundation.

\end{document}